\documentclass[prd,letterpaper,twocolumn,preprintnumbers,nofootinbib]{revtex4-2}
\usepackage[a4paper, hdivide={1.91cm,,1.165cm}, vdivide={1.83cm,,2.9cm}]{geometry}
\usepackage{subfigure}
\usepackage{slashed}
\usepackage{amsmath,amssymb}
\usepackage{graphicx}
\usepackage{units}
\usepackage{bbold}
\usepackage{xcolor}
\usepackage{dsfont}
\usepackage[hyperfootnotes=false,colorlinks,citecolor=blue]{hyperref}
\usepackage{comment}
\usepackage[normalem]{ulem}     

\newcommand{\beq}{\begin{equation}}
\newcommand{\eeq}{\end{equation}}
\newcommand{\bea}{\begin{eqnarray}}
\newcommand{\ena}{\end{eqnarray}}

\newcommand{\ii}{{\rm i}}

\def \L {\mathcal{L}} 

\def \epsilon {\varepsilon}

\newcommand{\hc}{\ensuremath{\text{H.c.}}}
\newcommand{\BR}{\ensuremath{\text{BR}}}

\newcommand{\re}{\ensuremath{\text{Re}}}
\newcommand{\im}{\ensuremath{\text{Im}}}

\newcommand{\matrixx}[1]{\begin{pmatrix} #1 \end{pmatrix}} 

\allowdisplaybreaks

\begin{document}

\preprint{CETUP-2023-005}

\title{Neutrino mass models at \texorpdfstring{$\mu$TRISTAN}{muTRISTAN}}

\author{P. S. Bhupal Dev}
\email[E-mail: ]{bdev@wustl.edu}
\thanks{ORCID: \href{https://orcid.org/0000-0003-4655-2866}{0000-0003-4655-2866}.}
\affiliation{Department of Physics and McDonnell Center for the Space Sciences, Washington University, St.~Louis, Missouri 63130, USA}

\author{Julian Heeck}
\email[E-mail: ]{heeck@virginia.edu}
\thanks{ORCID: \href{https://orcid.org/0000-0003-2653-5962}{0000-0003-2653-5962}.}
\affiliation{Department of Physics, University of Virginia,
Charlottesville, Virginia 22904-4714, USA}

\author{Anil Thapa}
\email[E-mail: ]{wtd8kz@virginia.edu}
\thanks{ORCID: \href{https://orcid.org/0000-0003-4471-2336}{0000-0003-4471-2336}.}
\affiliation{Department of Physics, University of Virginia,
Charlottesville, Virginia 22904-4714, USA}

\begin{abstract}
We study the prospects of probing neutrino mass models at the newly proposed antimuon collider $\mu$TRISTAN, involving $\mu^+e^-$ scattering at $\sqrt{s}= \unit[346]{GeV}$ and $\mu^+\mu^+$ scattering at $\sqrt{s}= \unit[2]{TeV}$.  We show that  $\mu$TRISTAN is uniquely sensitive to leptophilic neutral and doubly-charged scalars naturally occurring in various neutrino mass models, such as Zee, Zee--Babu, cocktail, and type-II seesaw models,  over a wide range of mass and coupling values, well beyond the current experimental constraints. It also allows for the possibility to correlate the collider signals with neutrino mixing parameters and charged lepton flavor violating observables.
\end{abstract}

\maketitle

\section{Introduction}
\label{sec:intro}

The origin of neutrino mass and mixing remains one of the important open questions in fundamental physics~\cite{Mohapatra:2006gs, deGouvea:2016qpx}. It clearly requires the introduction of new particles beyond the particle content of the Standard Model (SM).  
Qualitatively, we can expect these new particles to induce novel experimental signatures, such as lepton number violation (LNV) and charged lepton flavor violation (LFV), which are either forbidden or highly suppressed in the SM. 
Arguably, the cleanest method to identify the new particle(s) would be via their direct production at a high-energy collider. By studying the subsequent decays of these new particles to SM particles, preferably involving LNV and/or LFV to reduce SM background, one might be able to pinpoint the underlying neutrino mass model. A summary of existing collider constraints on various neutrino mass models can be found in Refs.~\cite{Deppisch:2015qwa, Cai:2017mow}. Similarly, a summary of the LFV constraints can be found in Refs.~\cite{Calibbi:2017uvl, Davidson:2022jai}.    

All past and current high-energy colliders constructed so far~\cite{Shiltsev:2012zzc} involve electron or proton beams and are therefore particularly sensitive to new particles that couple to electrons or quarks.
An entirely new class of couplings could be probed using muon colliders, originally proposed long ago~\cite{Budker:1969cd}. The main advantage is that leptons provide a  much cleaner collision environment than hadrons, and muon beams suffer less synchrotron radiation loss than electron beams, thus making muon colliders capable of reaching higher center-of-mass energies with a reasonable-size circular ring design~\cite{Ankenbrandt:1999cta, NeutrinoFactory:2002azy}. They have gained considerable attention in recent years~\cite{Delahaye:2019omf, Long:2020wfp, AlAli:2021let, MuonCollider:2022xlm, Accettura:2023ked}, as novel muon cooling techniques are now available~\cite{MICE:2019jkl}, and other technical difficulties related to the muon lifetime and radiation seem solvable~\cite{Accettura:2023ked}, making muon colliders an increasingly realistic and desirable option. 
Most work has been done in the context of future $\mu^+\mu^-$ colliders~\cite{Aime:2022flm}, which would mimic LEP~\cite{Hubner:2004qg} and could reach a center of mass energy of  \unit[10]{TeV} or more. 

Here, we will focus on a different experimental setup, 
$\mu$TRISTAN~\cite{Hamada:2022mua}, which is a proposed high-energy lepton collider using the ultra-cold antimuon technology developed at J-PARC~\cite{Abe:2019thb}. It can run in the $\mu^+ e^-$ mode with $\sqrt{s} = 346$ GeV, and later, in the $\mu^+ \mu^+$ mode~\cite{Heusch:1995yw} with $\sqrt{s}=2$ TeV or higher.
It can serve as a Higgs factory and do precision physics~\cite{Hamada:2022uyn}. Other new physics studies for the $\mu^+ e^-$ and $\mu^+ \mu^+$ collider options can be found in Refs.~\cite{Bossi:2020yne, Lu:2020dkx, Lichtenstein:2023iut} and \cite{Das:2022mmh, Yang:2023ojm, Lichtenstein:2023iut, Fridell:2023gjx}, respectively.   
As we will show in this article, the unique initial states of $\mu$TRISTAN make it especially sensitive to neutrino mass models involving leptophilic neutral and/or doubly-charged scalars, allowing for direct production and study of these new scalars in regions of parameter space otherwise untestable. We take examples from both tree- and loop-level neutrino mass models. Specifically, we use the Zee model~\cite{Zee:1980ai}, Zee--Babu model~\cite{Zee:1985id, Babu:1988ki}, cocktail model~\cite{Gustafsson:2012vj}, and type-II seesaw model~\cite{Konetschny:1977bn,Magg:1980ut,Schechter:1980gr,Cheng:1980qt,Mohapatra:1980yp} as concrete examples, and we consider the cleanest final states (with the least SM background), i.e., the LFV channels $\mu^+ e^- \to \ell_\alpha^+\ell_\beta^-$ and $\mu^+\mu^+\to \ell_\alpha^+\ell_\beta^+$ mediated by the scalars, as well as the associated production of scalars with a photon or $Z$ boson.\footnote{All models under consideration also generate LNV signatures, such as $\mu^+ \ell_\alpha^\pm\to W^+ W^\pm$, but since these are typically suppressed by a product of many couplings or even the neutrino mass, we will focus on LFV processes.} We show that $\mu$TRISTAN can provide unprecedented sensitivity  well beyond existing constraints and complementary to  future low-energy LFV searches.

The rest of this article is organized as follows: in Sec.~\ref{sec:muTRISTAN} we briefly describe the details of the $\mu$TRISTAN collider. In Sec.~\ref{sec:models} we go through several neutrino mass models (both radiative and tree-level), derive $\mu$TRISTAN's sensitivity and compare to other LFV observables, notably lepton flavor violation. We conclude in Sec.~\ref{sec:conclusions}.

\section{\texorpdfstring{$\mu$TRISTAN}{muTRISTAN}}
\label{sec:muTRISTAN}
The ultra-cold antimuon technology developed for the muon anomalous magnetic moment and electric dipole moment experiment at J-PARC~\cite{Abe:2019thb} uses laser ionization of muonium atoms to provide a low-emittance $\mu^+$ beam, which can be re-accelerated to high energies~\cite{Kondo:2018rzx}. Allowing a \unit[1]{TeV} $\mu^+$ beam to collide with a high-intensity $e^-$ beam at the TRISTAN (Transposable Ring Intersecting Storage Accelerators in Nippon~\cite{KEKTsukuba:1987ahw}) energy of \unit[30]{GeV} in a storage ring of the same size as TRISTAN (\unit[3]{km} circumference), one can realize the $\mu^+e^-$ mode of $\mu$TRISTAN with a center-of-mass energy $\sqrt s=\unit[346]{GeV}$.\footnote{A larger storage ring allows for higher-energy collisions. One can reach $\sqrt s=\unit[775]{GeV}$ with \unit[50]{GeV} electrons and \unit[3]{TeV} muons.} Taking into account muon decay, the deliverable instantaneous luminosity for a single detector at any collision point in the storage ring is estimated as $4.6\times 10^{33}\ {\rm cm}^{-2}\ {\rm s}^{-1}$~\cite{Hamada:2022uyn}, which translates to an integrated luminosity of $100\ {\rm fb}^{-1}\ {\rm year}^{-1}$. 

Using the same 3 km storage ring and 1 TeV $\mu^+$ beams, one can also consider a $\mu^+\mu^+$ collider~\cite{Heusch:1995yw} with $\sqrt s=\unit[2]{TeV}$ (or 6 TeV for the larger ring option). The beam intensity will be lower than in the $\mu^+e^-$ mode due to both muons decaying in the storage ring. The instantaneous luminosity is estimated as $5.7\times 10^{32}\ {\rm cm}^{-2}\ {\rm s}^{-1}$~\cite{Hamada:2022uyn}, which translates to an integrated luminosity of $12~{\rm fb}^{-1}~{\rm year}^{-1}$. 

The precise luminosity numbers depend on various efficiencies for the muon production, as well as the detailed designs of the muon accelerator and storage ring. For instance, a higher luminosity is, in principle, achievable with better focusing of the $e^-$ beam (compared to the $\mu^+$ beam~\cite{Abe:2019thb}), following the SuperKEKB design~\cite{Belle-II:2010dht}. We will use the numbers given above from Ref.~\cite{Hamada:2022uyn} as realistic but conservative order-of-magnitude estimates to work with.  
Assuming negligible SM background for the LFV signals we study below, the above-mentioned luminosities correspond to a minimum signal cross section of 0.09 (0.75) fb in the $\mu^+e^-$ ($\mu^+\mu^+$) mode in order to achieve $3\sigma$ sensitivity with 1 year runtime. To be conservative, we will use a signal cross section of 0.1 (1) fb in the $\mu^+e^-$ ($\mu^+\mu^+$) mode to derive our sensitivity limits. These limits can be easily scaled for a longer runtime. For instance, 10 years of runtime with 1 ab$^{-1}$ integrated luminosity can achieve the same level of sensitivity with a signal cross section ten times smaller, thus being capable of probing a larger model parameter space than what is shown here. 

Since the details of the $\mu$TRISTAN detector design and acceptance efficiencies are currently unknown, we will only impose basic trigger-level cuts
 on the transverse momenta and pseudorapidity of the outgoing leptons and photons, i.e., the default {\tt MadGraph5} cuts $p_T^{\ell,\gamma}>10$ GeV and $|\eta^{\ell,\gamma}|<2.5$~\cite{Alwall:2014hca}  while calculating the cross sections in the $\mu^+\mu^+$ option. For the asymmetric beams in the $\mu^+e^-$ option, we only keep the trigger-level $p_T$ cuts and remove the $\eta$ cuts because the final state particles are boosted in the $\mu^+$ direction; the detector should be designed to cover the small-angle region from the beam direction.

We will use unpolarized beams for both $\mu^+e^-$ and $\mu^+\mu^+$ modes to derive our sensitivity limits. Although the surface antimuons produced by the $\pi^+$ decay are 100\% polarized due to the $V-A$ nature of the weak interaction, the final polarization of the antimuon beam depends on a detailed understanding of the beam emittance under the applied magnetic field, which in some cases can reduce the polarization down to 25\%~\cite{Hamada:2022uyn}. Similarly, the beam polarization option for the $e^-$ beam is still under discussion for the SuperKEKB upgrade~\cite{Roney:2021pwz}. Including realistic beam polarization effects could modify our cross sections by a factor of few due to the chiral nature of the scalar couplings. 

\section{Neutrino mass models with leptophilic scalars}
\label{sec:models}
The leptonic initial states and clean environment at $\mu$TRISTAN provide an unprecedented opportunity to directly probe heavy leptophilic particles with possible LFV interactions. We will mainly focus on the leptophilic neutral and doubly-charged scalars that arise in well-known neutrino mass models, both tree-level and radiative, such as the Zee model~\cite{Zee:1980ai}, Zee--Babu model~\cite{Zee:1985id, Babu:1988ki}, cocktail model~\cite{Gustafsson:2012vj}, and type-II seesaw model~\cite{Konetschny:1977bn,Magg:1980ut,Schechter:1980gr,Cheng:1980qt,Mohapatra:1980yp}. If kinematically allowed, a neutral scalar $H$ with sizable LFV coupling $e\mu$ can be resonantly produced in $\mu^+e^-$ collisions either by itself or in association with a photon or $Z$ boson, as shown in Fig.~\ref{fig:neutralhiggs}(a) and (b) respectively, thus providing unparalleled sensitivity to the LFV scalar sector. Even for $m_H>\sqrt s$, the dilepton channels $\mu^+e^-\to \ell_\alpha^+\ell_\beta^-$ and $\mu^+\mu^+\to \ell_\alpha^+\ell_\beta^+$, shown in Fig.~\ref{fig:neutralhiggs}(c) and (d), respectively, are sensitive to the LFV couplings of $H$  and give rise to a contact-interaction-type bound on the scalar parameter space. Similarly, a doubly-charged scalar can be resonantly produced at a $\mu^+\mu^+$ collider, either by itself or in association with a photon or $Z$ boson (see Fig.~\ref{fig:doublycharged}). The higher center-of-mass energy of the $\mu^+\mu^+$ option at $\mu$TRISTAN allows us to probe doubly-charged scalars beyond the current LHC constraints~\cite{ATLAS:2022pbd}. We only focus on the LFV final states, as they are free from the SM background (modulo lepton misidentification, whose rate is negligible at lepton colliders~\cite{CEPCStudyGroup:2018ghi, Bartosik:2020xwr}). Also, we do not consider processes involving singly-charged scalars, as they necessarily involve neutrinos in the final state, making it harder to separate our signal from the SM background.

\subsection{Zee model}
In the Zee model~\cite{Zee:1980ai}, the SM scalar sector with one Higgs doublet $H_1$ is extended by adding a second Higgs doublet $H_2$ and an $SU(2)_L$-singlet charged scalar $\eta^+$. The relevant Lagrangian terms are given by
\begin{align}
\L \supset \mu H_1 H_2 \eta^- - f\bar{L}^c  L  \eta^+ - \tilde{Y}\bar{\ell} L \tilde{H}_1- Y \bar{\ell} L \tilde{H}_2 +\hc \, ,
\label{eq:lag}
\end{align}
where the superscript $c$ stands for charge conjugate and $\tilde{H}_a\equiv i\sigma_2H_a^\star$ ($a=1,2$, $\sigma_2$ is the second Pauli matrix). We have suppressed the flavor and $SU(2)_L$ indices. Note that the Yukawa coupling matrix $f$ is anti-symmetric in flavor space, while $Y$ is an arbitrary complex coupling matrix. 
We go to the Higgs basis~\cite{Georgi:1978ri, Davidson:2005cw}, where only $H_1$ acquires a vacuum expectation value, $\langle H_1\rangle 
\equiv v/\sqrt{2} \simeq \unit[174]{GeV}$, and the charged leptons obtain a diagonal mass matrix
$M_\ell = \tilde{Y} v/\sqrt{2}$.
We work in the alignment limit~\cite{Gunion:2002zf}, as preferred by the LHC Higgs data~\cite{Eberhardt:2020dat}, where the neutral scalars of $H_2$ (the CP-even $H$ and the CP-odd $A$) do not mix with the neutral Higgs contained in $H_1$ that can be identified as the SM Higgs boson. 
The $\mu$ term in the Lagrangian~\eqref{eq:lag} will induce a mixing of $\eta^+$ with the charged scalar contained in $H_2$ upon electroweak symmetry breaking; we denote the mixing angle by $\phi$ and the two mass eigenstates by $h^+$ and $H^+$, see Refs.~\cite{Babu:2019mfe, Barman:2021xeq} for details.

The simultaneous presence of $f$, $Y$, and $\mu$ breaks lepton number by two units and leads to a one-loop Majorana neutrino mass matrix
\begin{align}
M^\nu = \kappa \left( f M_\ell Y + Y^T M_\ell f^T\right),
\label{eq:Mnu}
\end{align}
with prefactor $\kappa\equiv (16\pi^2)^{-1} \sin 2\phi \log(m_{h^+}^2/m_{H^+}^2)$.
This matrix is manifestly symmetric and can be diagonalized as usual via
\begin{align}
    M^\nu = U \,\text{diag}(m_1,m_2,m_3)\, U^T\,,
\end{align}
where $U$ is the unitary Pontecorvo--Maki--Nakagawa--Sakata matrix and $m_j$ the neutrino masses.
Through neutrino oscillations we have obtained information about the mass splittings and the three mixing angles in $U$. The overall neutrino mass scale, ordering, and CP phases are unknown, although their ranges are partially restricted~\cite{Esteban:2020cvm}.

\begin{figure}[t!]
    \centering
    \subfigure[]{
    \includegraphics[scale=0.4]{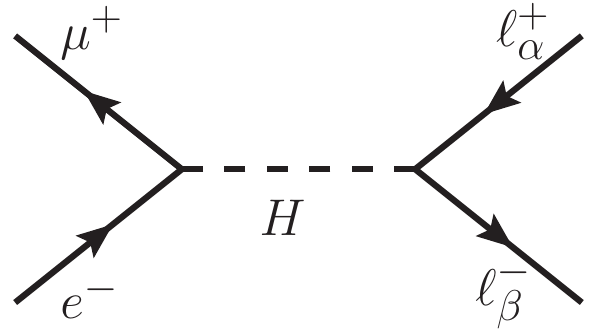}}
    \hspace{0.2cm}
    \subfigure[]{
    \includegraphics[scale=0.4]{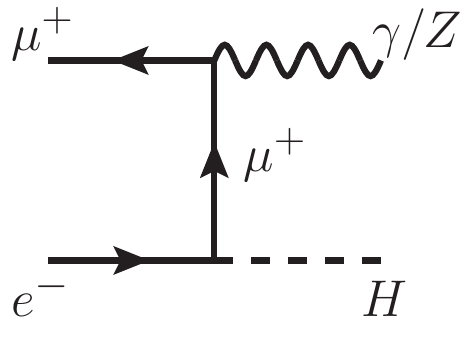}}\\ 
    \subfigure[]{
    \includegraphics[scale=0.4]{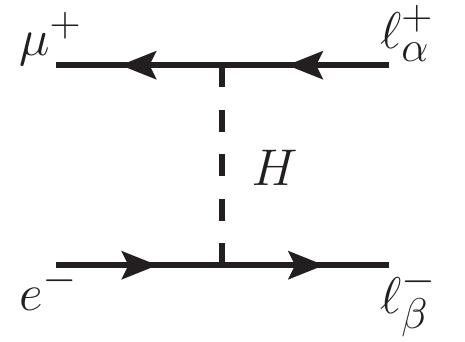}}
    \hspace{0.2cm}
    \subfigure[]{
    \includegraphics[scale=0.4]{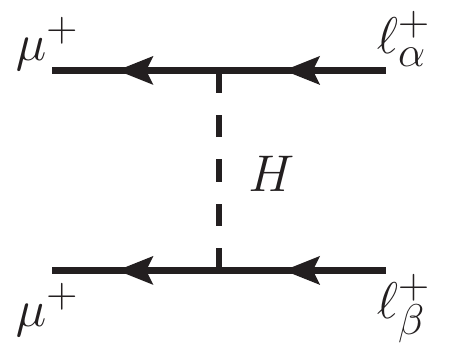}}
    \caption{Relevant Feynman diagrams for the processes involving the neutral scalar $H$ in the Zee model  at $\mu$TRISTAN.}
    \label{fig:neutralhiggs}
\end{figure}

With the parametrization of Refs.~\cite{Machado:2017flo,Heeck:2023iqc} we can express $Y$ in terms of $M^\nu$ and $f$. The $\mu^+ e^-$ run  of $\mu$TRISTAN will be uniquely sensitive to $Y_{e\mu}$ and $Y_{\mu e}$, see Fig.~\ref{fig:neutralhiggs}(a)-(c), so we investigate $Y$ textures where one of these entries is non-vanishing, which is hardly a restriction. The simultaneous presence of $Y_{e\mu}$ and $Y_{e e}$ (or $Y_{\mu\mu}$) however would induce large LFV amplitudes, e.g.~$\mu\to e \gamma$ and $\mu\to 3 e$~\cite{Lavoura:2003xp, He:2011hs, Herrero-Garcia:2017xdu, Cai:2017jrq, Crivellin:2015hha}, leaving little parameter space for $\mu$TRISTAN to probe. To evade LFV constraints  and simplify our analysis, we will set as many $Y$ entries to zero as possible, leading to the four benchmark textures
\begin{align}
Y_{A_1} & \propto 
   \matrixx{ 
   0 &  1  & 0\\
    0 & 0 & - \frac{2 m_e}{m_\mu} \frac{M^\nu_{e\tau}}{M^\nu_{\mu\mu}} \\
    0 & 0  & 0 } \sim 
    \matrixx{ 
   0 &  1  & 0\\
    0 & 0 & 0.0035 \\
    0 & 0  & 0 },\\
Y_{B_2} & \propto 
   \matrixx{ 
   0 &  1  & 0\\
    -\frac{m_e}{m_\mu} \frac{M^\nu_{e e}}{M^\nu_{\mu\mu}} & 0 & 0 \\
    0 & 0  & 0 } \sim
     \matrixx{ 
   0 &  1  & 0\\
    0.013  & 0 & 0 \\
    0 & 0  & 0 },\\
Y_{B_3} & \propto 
   \matrixx{ 
   0 &  0  & 1\\
    -\frac{m_e}{2 m_\mu} \frac{M^\nu_{e e}}{M^\nu_{\mu\tau}} & 0 & 0 \\
    0 & 0  & 0 }\sim
     \matrixx{ 
   0 &  0  & 1 \\
    0.0023 & 0 & 0 \\
    0 & 0  & 0 },\\
Y_{B_4} & \propto 
   \matrixx{ 
   0 &  1 & 0\\
    0 & 0 & 0 \\
    -\frac{m_e}{2 m_\tau} \frac{M^\nu_{e e}}{M^\nu_{\mu \tau}} & 0  & 0 } \sim
    \matrixx{ 
   0 &  1 & 0\\
    0 & 0 & 0 \\
    0.00013 & 0  & 0 }.
\end{align}
All these $Y$ textures lead to viable two-zero textures in $M^\nu$~\cite{Alcaide:2018vni}, indicated by their common name as a subscript, following the nomenclature of Ref.~\cite{Frampton:2002yf}.
The $M^\nu$ two-zero textures predict the unknown parameters in the neutrino sector,  i.e., the lightest neutrino mass and the three phases.
We show in Tab.~\ref{tab:texture_zeros} the predictions for the sum of neutrinos masses $\sum_j m_j$ (testable via cosmology~\cite{Dvorkin:2019jgs}), the effective mass parameter for neutrinoless double beta decay $\langle m_{\beta\beta}\rangle =\sum_i U^2_{ei}m_i$ (testable in the next-generation experiments~\cite{Adams:2022jwx}), and the Dirac CP phase (testable in  neutrino oscillation experiments~\cite{Hyper-KamiokandeProto-:2015xww, DUNE:2020jqi}). Notice that  the $\sum m_\nu$ predictions of the $B$ textures are already in tension~\cite{Meloni:2014yea} with limits from cosmology, $\sum m_\nu < \unit[0.12]{eV}$~\cite{Planck:2018vyg},\footnote{ Even stronger limits have been obtained in Refs.~\cite{Palanque-Delabrouille:2019iyz,diValentino:2022njd}, while mild indications of a nonzero sum of neutrino masses (in tension with the stringent Planck limits) was suggested in Ref.~\cite{DiValentino:2021imh}.} but perfectly in line with laboratory constraints~\cite{KATRIN:2021uub}. 

\begin{table}[tb]
\begin{tabular}{c|c|c|c|c}
name & texture zeros & $\sum_j m_j/\unit{eV}$ & $\langle m_{\beta\beta}\rangle/\unit{eV}$ & $\delta_{\rm CP}/{}^\circ$ \\
\hline
$A_1$ &$M_{ee}$, $M_{\mu e}$ &  $0.062$--$0.071$ & $0$ & $44$--$341$ \\
$B_2$ &$M_{\tau\tau}$, $M_{\mu e}$ &  $>0.13$ & $>0.036$ & $85$-$90$ $\wedge$ $270$-$275$\\
$B_3$ &$M_{\mu\mu}$, $M_{\mu e}$ & $>0.16$ & $>0.047$ & $87$-$90$ $\wedge$ $270$-$273$\\
$B_4$ &$M_{e\tau}$, $M_{\tau\tau}$ &  $>0.14$ & $>0.039$ & $90$-$94$ $\wedge$ $266$-$270$
\end{tabular}
\caption{Predictions for the sum of neutrino masses $\sum_j m_j$, the effective $0\nu\beta\beta$ Majorana neutrino mass $\langle m_{\beta\beta}\rangle$, and the Dirac CP phase $\delta_{\rm CP}$ from the texture zeros employed in the Zee model, using the $3\sigma$ normal-ordering ranges for the oscillation parameters from {\tt NuFit 5.2}~\cite{Esteban:2020cvm}.
}
\label{tab:texture_zeros}
\end{table}

The many zeros in these four $Y$ benchmarks ensure highly suppressed LFV. 
Indeed, neither of them give rise to the most stringent LFV modes, $\mu\to e\gamma$ and $\mu\to 3e$, despite the non-zero $e\mu$ entry in $Y$. However, all cases induce muonium--antimuonium oscillation~\cite{Pontecorvo:1957cp,  Jentschura:1997tv, Clark:2003tv, Fukuyama:2021iyw} through those $e\mu$ entries, which will turn out to be an important constraint. In addition, all textures except for $Y_{B_2}$ also give rise to LFV tauon decays. 
Furthermore, all textures
contribute to $(g-2)_\mu$, although the $2\sigma$-preferred region turns out to be already excluded by the muonium constraint.

The overall scale of $Y$ is degenerate with $f$  and $\kappa$ from Eq.~\eqref{eq:Mnu} and can effectively be adjusted at will. The $e\mu$ entry of $Y$ is then a free parameter, subject only to perturbative unitarity constraints.
The second non-zero entry of $Y$ is not free, however, but rather predicted by lepton masses and neutrino mass matrix entries. The latter are essentially predicted due to the two-zero textures in $M^\nu$, allowing us to predict the $Y$ entries, as already shown above.
For $A_1$, $B_2$, and $B_4$, we find a large $e\mu$ entry in $Y$ that drives the $H$ production at $\mu$TRISTAN, plus a suppressed second $Y$ entry that induces LFV. For $B_3$, the $e\tau$ entry dominates and $\mu$TRISTAN's reach is severely limited by tau LFV.
Notice that we are focusing on such extreme textures just for the sake of illustration to emphasize $\mu$TRISTAN's complementarity to other experimental probes. 

\begin{figure}[!t]
\includegraphics[width=0.48\textwidth]{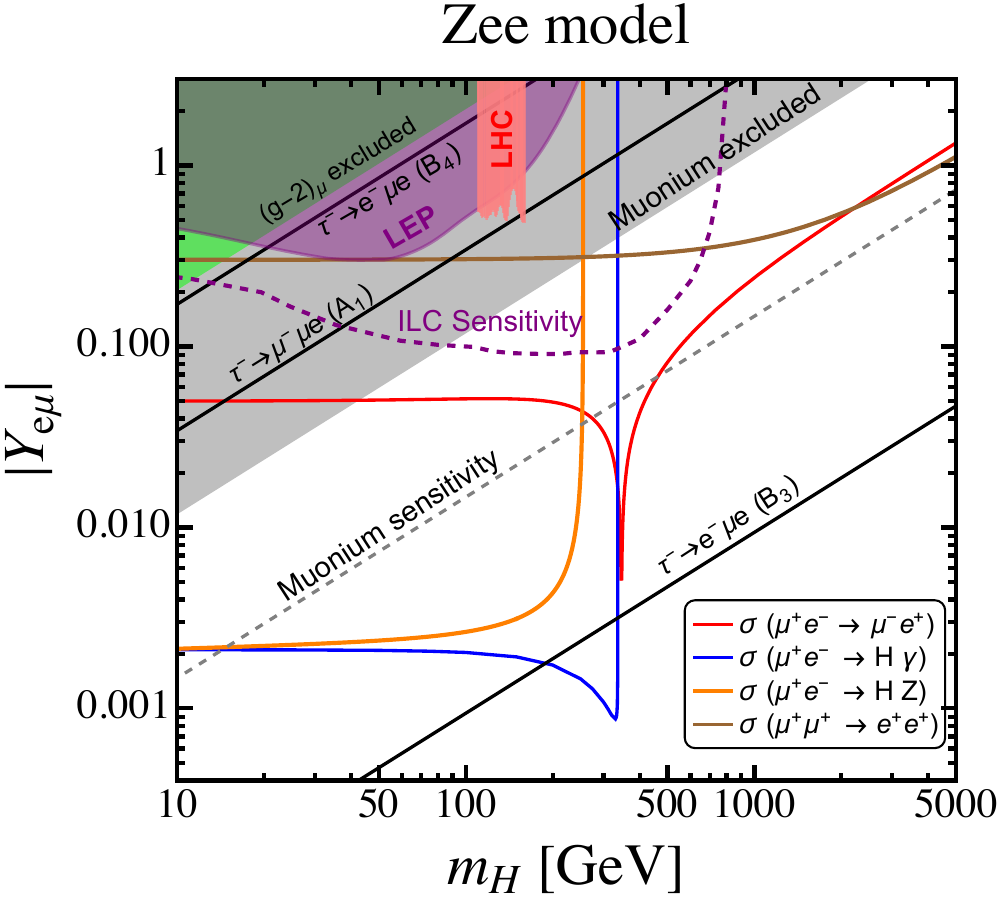}
    \caption{$\mu$TRISTAN sensitivity to the Zee model parameter space for various channels as shown in Fig.~\ref{fig:neutralhiggs}.  The shaded regions are excluded: Purple (pink) shaded from LEP (LHC) dilepton data, green shaded from $(g-2)_\mu$, and gray shaded from muonium oscillation. The future muonium (ILC) sensitivity is shown by the black (purple) dashed line (curve). The solid black lines show the $\tau$ LFV constraints for different $Y$ textures ($A_1, B_3, B_4$). 
   } 
    \label{fig:Zeeplot}
\end{figure}

Assuming $H$ to be the lightest scalar, the textures $Y_{A_1}$, $Y_{B_3}$, and $Y_{B_4}$ lead to $\tau^-\to \mu^-\mu^\pm e^\mp$, $\tau^-\to e^-\mu^\pm e^\mp$, and $\tau^-\to e^-e^\pm\mu^\mp$, respectively, 
which give limits of order $|Y_{\tau\alpha} Y_{\beta\delta}|< (m_H/\unit[5]{TeV})^2$, as shown by the solid black lines in Fig.~\ref{fig:Zeeplot}. For all textures except $B_3$ these are very suppressed by the small $Y_{\tau\alpha}$ entry. For those textures, as well as for the $Y_{B_2}$ texture which does not give rise to tau (or muon) LFV decay, the most important LFV process is the $|\Delta L_\mu| =|\Delta L_e|=2$ conversion of muonium ($M = e^-\mu^+$) to  antimuonium ($\bar{M}= e^+\mu^-$)~\cite{Pontecorvo:1957cp,  Jentschura:1997tv, Clark:2003tv, Fukuyama:2021iyw}, which only requires the $Y_{e\mu}$ entry we are interested in for $\mu$TRISTAN.
The conversion probability is currently limited to $P(M\leftrightarrow\bar{M}) < 8.3 \times 10^{-11}$ at $90\%$ C.L. by the MACS experiment at PSI~\cite{Willmann:1998gd}, while a sensitivity at the level of $\mathcal{O}(10^{-14})$ is expected in the future by the proposed MACE experiment~\cite{Bai:2022sxq}. The current MACS limit sets stringent constraints on the Yukawa couplings $Y_{e\mu}$ and $Y_{\mu e}$:
\begin{align}
 |Y_{e\mu,\mu e}| < \frac{m_H}{\unit[0.85]{TeV}} \,.
 \label{eq:muonium}
\end{align}
This is the most important limit for $\mu$TRISTAN, as shown in Fig.~\ref{fig:Zeeplot} by the gray-shaded region (current) and black dotted line (future).

The muonium limit can be significantly weakened due to destructive interference in the $M-\bar{M}$ amplitude~\cite{Afik:2023vyl} if we choose $m_A\simeq m_H$, which renders even the future MACE projection insensitive to our parameter space of interest. However, for $m_H\simeq m_A\ll m_{H^+}$,  we would generate large oblique parameters due to custodial symmetry breaking~\cite{Peskin:1990zt, Peskin:1991sw}; this puts an upper limit on the mass splitting between the neutral and charged scalars in the Zee model~\cite{Heeck:2023iqc, Afik:2023vyl}. On the other hand, the leptophilic charged scalars in this model are constrained from slepton searches at the LHC because the slepton decay $\tilde{\ell}^+\to \ell^+\tilde{\chi}^0$ mimicks a charged scalar decay $H^+\to \ell^+\nu$ in the massless neutralino limit. The current LHC bound is $m_{H^+}>425$~GeV at 90\% CL~\cite{ATLAS:2019lff} for ${\rm BR}(H^+\to \mu^+\nu_e)=1$. To
evade the muonium bound while satisfying the global electroweak precision constraint~\cite{Lu:2022bgw, Asadi:2022xiy}, we then require $m_H\simeq m_A\gtrsim \unit[320]{GeV}$, making direct $H$ production in $\mu$TRISTAN's $\mu^+ e^-$ mode difficult. To extend our analysis to lighter $H$, we therefore assume the scalar hierarchy $m_H\ll m_A\simeq m_{H^+}$, subject to the muonium constraint from Eq.~\eqref{eq:muonium}.\footnote{Note that our results are symmetric under $m_H \leftrightarrow m_A$; we simply choose $H$ to be the lighter one for concreteness.} Moreover, to set the scale of neutrino masses, we choose the $f$ couplings to be much smaller than $Y$ and can hence neglect the $\eta^\pm$-mediated processes at $\mu$TRISTAN entirely.

Having established our benchmark scenarios and relevant LFV signatures, we can study this region of the Zee-model parameter space at $\mu$TRISTAN. The relevant Feynman diagrams and processes are shown in Fig.~\ref{fig:neutralhiggs}. 
Away from the $s$-channel resonance at $\sqrt{s}\sim m_H$, the dilepton cross section takes on the simple form
\begin{align}
    \sigma (\mu^+ e^-\to \mu^- e^+) \simeq \frac{|Y_{e\mu}|^4}{64\pi s} \begin{cases}
        1\,,  &  m_H\ll \sqrt{s}\,,\\
        \frac{s^2}{12 m_H^4} \,, &  m_H\gg \sqrt{s}\, .
    \end{cases}
\end{align}
This was numerically verified in {\tt MadGraph5\_aMC@NLO}~\cite{Alwall:2014hca}
 using the general 2HDM {\tt FeynRules} model file~\cite{Alloul:2013bka}. The exact analytic expression for the cross section is not very illuminating, and therefore, we do not show it here. 
We demand this cross section to be of order $\unit[0.1]{fb}$ (after applying the cuts specified in Sec.~\ref{sec:muTRISTAN}) for a discovery, since this flavor-violating channel is background-free. The textures $A_1$, $B_2$, and $B_4$ dominantly induce this channel.\footnote{The $B_3$ texture dominantly gives  $\mu^+ e^-\to \tau e$.} 
We show the $\mu$TRISTAN reach of this process $\mu^+ e^-\to  \mu^- e^+$ in Fig.~\ref{fig:Zeeplot} (solid red curve), after applying the basic trigger cuts. 
We find that the $\mu$TRISTAN sensitivity surpasses the current limit from muonium conversion for $m_H > \unit[50]{GeV}$. The $B_3$ texture is the only one that is already too constrained by tau LFV to give large $\sigma (\mu^+ e^-\to \ell^+_\alpha \ell^-_\beta)$.
Future muonium data can cover almost the entire relevant parameter space for $\mu$TRISTAN's dilepton mode in the Zee model, offering confirmation potential in case of a discovery. 

In Fig.~\ref{fig:Zeeplot}, we also show the existing collider constraints from LEP $e^+e^-\to \mu^+\mu^-$ data (purple shaded)~\cite{OPAL:2003kcu, Electroweak:2003ram} and from LHC $pp\to e\mu$ data (pink shaded)~\cite{ATLAS:2019old, CMS:2023pte}.\footnote{As noted in Ref.~\cite{Afik:2023vyl}, the $3.8\sigma$ CMS excess in the $e\mu$ channel~\cite{CMS:2023pte} can be explained by $H$ using lepton PDF, but only for $m_H\simeq m_A$ to avoid the muonium limit.} The future ILC sensitivity from $e^+e^-\to  \mu^+\mu^-H$ is also shown by the pink dashed curve~\cite{Dev:2017ftk, Dev:2018vpr, Barman:2021xeq} for comparison with the $\mu$TRISTAN sensitivity. The green-shaded region is excluded by demanding the $H$ contribution to $(g-2)_\mu$ not to exceed $5\sigma$ deviation between the world average of the SM prediction~\cite{Aoyama:2020ynm} and the experimental value~\cite{Muong-2:2023cdq}.\footnote{Taking the BMW result~\cite{Borsanyi:2020mff} instead of the world average~\cite{Aoyama:2020ynm} for the SM prediction does not make much difference to our allowed parameter space, which is dominated by the muonium limit.}

For the associated production of $H$ with a photon or a $Z$ boson (cf.~Fig.~\ref{fig:neutralhiggs}(b)), the cross sections for small $m_H\ll \sqrt s$ take the form
\begin{align}
   & \sigma (\mu^+ e^-\to H\gamma) \simeq \frac{\alpha_\text{EM} |Y_{e\mu}|^2}{8 s}\,\log\left(\frac{s}{m_e m_\mu}\right) ,\\
&\sigma (\mu^+ e^-\to H Z) \simeq  \frac{\alpha_\text{EM} |Y_{e\mu}|^2 \, (s-m_Z^2)}{32 s_w^2 c_w^2 s^2 }\left[ \frac{s}{4 m_Z^2}  \right.\\
& \left.-  (1-2 s_w^2 + 4 s_w^4) -  (1-4 s_w^2 + 8 s_w^4) \log\left(\frac{m_H m_Z}{s-m_Z^2}\right)\right], \nonumber
\end{align}
where $\alpha_{\rm EM}$ is the electromagnetic fine-structure constant, and $s_w\equiv \sin\theta_w$ ($c_w\equiv \cos\theta_w$) is the (co)sine of the weak mixing angle. 
These cross sections are typically larger than the dilepton channel but are open only for $m_H \lesssim \sqrt{s}$ for the photon case (or $\sqrt{s}-m_Z$ for the $Z$ case).
The photon cross section exhibits an infrared divergence for $\sqrt{s}\to m_H$ that is regulated by the cut $p_T^\gamma >\unit[10]{GeV}$, reducing the total cross section  compared to the analytical expression above. 
The $Z$ cross section is well behaved near the kinematic threshold but diverges for $m_H\to 0$, not of any concern for us. 
As can be seen in Fig.~\ref{fig:Zeeplot}, both modes are important for $\mu$TRISTAN and cover parameter space that cannot be probed with other colliders or LFV.\footnote{The only exception is texture $B_3$, for which only a tiny region survives the tau LFV bound.} 
The $H$ scalars subsequently decay promptly into $\mu^\pm e^\mp$, half of which being background free even without any momentum reconstruction.

The Zee model also makes predictions for $\mu$TRISTAN's $\mu^+\mu^+$ mode, as there are $t$-channel diagrams for $\mu^+\mu^+\to \ell^+ \ell'^+$ (cf.~Fig.~\ref{fig:neutralhiggs}(d)). All textures except $B_3$ induce the background free $\mu^+\mu^+\to e^+ e^+$, with testable allowed cross sections for $m_H>\unit[300]{GeV}$, as shown in Fig.~\ref{fig:Zeeplot} by the brown curve. We find that the $H$ sensitivity in this channel is worse than or comparable to the dilepton channel in the $\mu^+e^-$ mode, so it can only be used as a secondary channel for verifying any signal found in $\mu$TRISTAN's first run.

Before we move on to other neutrino mass models, let us briefly comment on the discrepancy in the muon magnetic moment~\cite{Muong-2:2023cdq}. While the status of the SM prediction is currently unclear, it is worthwhile to entertain the possibility that the discrepancy is real and a sign for new physics. The benchmark values taken above are incapable of explaining $(g-2)_\mu$ due to LFV constraints. A recent study~\cite{Heeck:2023iqc} has shown that the Zee model is in principle able to explain $(g-2)_\mu$, but this requires one of the following textures:
\begin{align}
Y =
   \matrixx{ 
   0 & 0 & 0\\
    0 & \times & \times \\
    0 & \times  & \times } \text{ or }
    \matrixx{ 
   \times & 0 & \times\\
    0 & \times & 0 \\
    \times & 0  & \times } .
\end{align}
The first (second) requires $M^\nu_{ee} = 0$ ($M^\nu_{\mu\mu} = 0$) and effectively conserves electron (muon) number, which makes it obvious that muon LFV is evaded, including muonium conversion. The first texture could only show up in $\mu$TRISTAN's $\mu^+\mu^+$ run via $\mu^+\mu^+\to \mu^+\tau^+$ or $\tau^+\tau^+$;
the second texture can give $\mu^+ e^- \to \mu^+ \tau^-$ in $\mu$TRISTAN's first run. A dedicated study of this scenario will be postponed until the $(g-2)_\mu$ anomaly is clarified.

Overall, we see that $\mu$TRISTAN could probe the Zee model in regions of parameter space that are inaccessible by other means. A exhaustive study of the Zee model at $\mu$TRISTAN goes beyond the scope of this work but the benchmarks discussed here indicate a very promising situation.

\subsection{Zee--Babu model}
In the Zee--Babu model~\cite{Zee:1985id,Babu:1988ki}, we extend the SM by two $SU(2)_L$-singlet scalars $h^+$ and $k^{++}$ with hypercharge $1$ and $2$, respectively, which have the following couplings relevant for neutrino masses:
\begin{align}
-\L \supset f\bar{L}^c  L h^+ + g\bar{\ell}^c  \ell\, k^{++} + \mu h^- h^- k^{++}+\hc
\end{align}
The matrix $g$ ($f$) is symmetric (antisymmetric) in flavor space.
Taken together, these couplings break lepton number and generate a Majorana neutrino mass matrix
\begin{align}
M^\nu \simeq 16\mu\, I(m_h,m_k)\, f M_\ell g^* M_\ell f\,,
\end{align}
where $I(m_h,m_k)$ is a two-loop function~\cite{Babu:2002uu, Nebot:2007bc}.
The antisymmetry of $f$ leads to $\det M^\nu = 0$ and thus predicts one massless neutrino.

\begin{figure}
    \centering
    \subfigure[]{\includegraphics[scale=0.4]{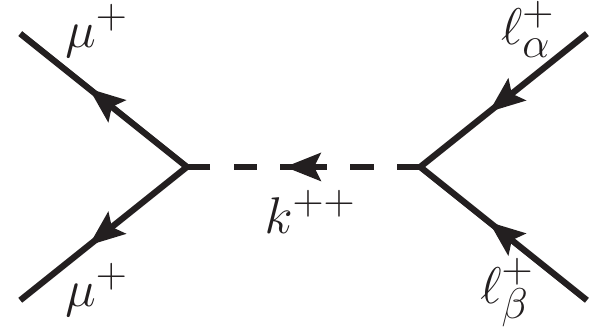}}
    \hspace{0,2cm}
    \subfigure[]{
    \includegraphics[scale=0.4]{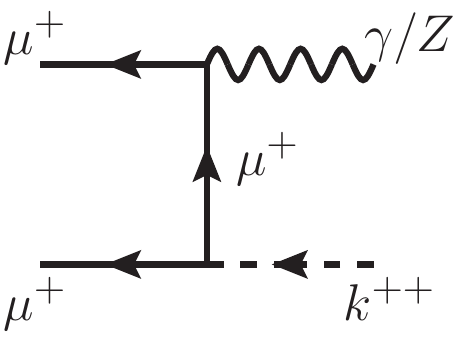}}
    \\ 
    \subfigure[]{
    \includegraphics[scale=0.4]{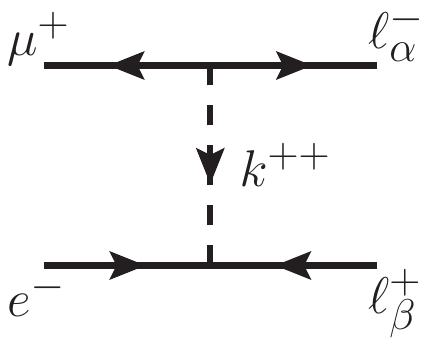}}
    \caption{Relevant Feynman diagrams for the doubly-charged scalars in the Zee--Babu, cocktail, and triplet seesaw models. 
    }
    \label{fig:doublycharged}
\end{figure}

\begin{figure}[!t]
    \centering
    \includegraphics[width=0.46\textwidth]{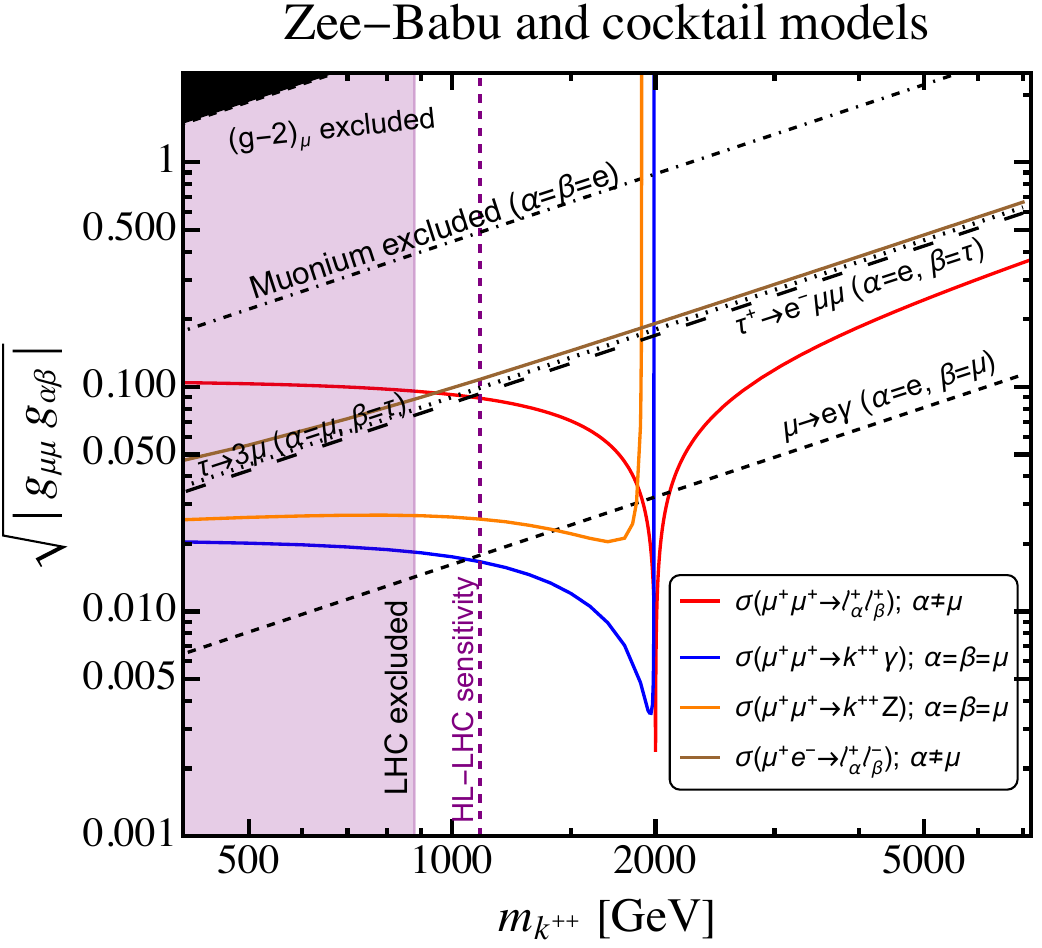}
    \caption{
    $\mu$TRISTAN sensitivity to the Zee--Babu and cocktail model parameter space for various channels as shown in Fig.~\ref{fig:doublycharged}.  The shaded purple region is excluded from LHC dilepton data \cite{ATLAS:2022pbd}, the dashed purple line shows the HL-LHC reach~\cite{Ruiz:2022sct}.
  The diagonal non-solid lines indicate LFV constraints on the coupling products $|g_{\mu\mu} g_{\alpha\beta}|$. For the Zee--Babu $g$ texture from Eq.~\eqref{eq:ZeeBabu_texture}, only $g_{\mu\mu}$ and $g_{\mu\tau}$ are relevant.
  For the cocktail-model texture from Eq.~\eqref{eq:Cocktail_texture}, mainly $g_{\mu\mu}$ and $g_{e\tau}$ are relevant. 
    }
    \label{fig:ZeeBabuplot}
\end{figure}

Similar to the Zee model, we can make the overall scale of $g$ as large as we want and compensate for that with a smaller $f$ matrix or $\mu$ coupling.
For simplicity we assume $h^+$ to be very heavy and the $f$ couplings to be small, effectively decoupling $h^+$. This leaves us with the doubly charged $k^{++}$ with coupling matrix $g$.
At $\mu$TRISTAN's $\mu^+\mu^+$ run, this $k^{++}$ leads to dilepton and associated production signatures as long as $g_{\mu\mu}\neq 0$, see Fig.~\ref{fig:doublycharged}(a)-(b). We show $\mu$TRISTAN's reach  and competing constraints in Fig.~\ref{fig:ZeeBabuplot}, having  computed the cross sections with {\tt MadGraph5\_aMC@NLO}~\cite{Alwall:2014hca} using the model file given in Ref.~\cite{Ruiz:2022sct}.

$\mu$TRISTAN  can easily probe a large region of parameter space as long as $g_{e\mu}$ is somewhat suppressed compared to $g_{\mu\mu}$ to evade the $\mu\to e \gamma$ constraint. This is hardly a restriction and we can even find $g$ textures that eliminate almost all LFV constraints, e.g.
\begin{align}
g^* &\propto \matrixx{
0 & 0 & 0 \\
0 & 1 & -\frac{m_\mu}{m_\tau}\, \frac{M^\nu_{\mu\tau}}{M^\nu_{\tau\tau}} \\
0 & -\frac{m_\mu}{m_\tau}\, \frac{M^\nu_{\mu\tau}}{M^\nu_{\tau\tau}} & \frac{m_\mu^2}{m_\tau^2}\, \frac{M^\nu_{\mu\mu}}{M^\nu_{\tau\tau}}
}
\sim\matrixx{
0 & 0 & 0 \\
0 & 1 & 0.1 \\
0 & 0.1 & 5 \times 10^{-3}} .
\label{eq:ZeeBabu_texture}
\end{align}
This structure does not lead to any $e$LFV. The only process we could worry about is $\tau\to 3\mu$, which is however not particularly stringent and could be further suppressed by tuning $|M^\nu_{\mu\tau}/M^\nu_{\tau\tau}|\ll1$. $\mu$TRISTAN has a large region of testable parameter space even without this tuning. Notice that the dominant $g_{\mu\mu}$ entry here leads to the dominant channels $\mu^+\mu^+\to \mu^+\mu^+$ and $\mu^+\mu^+\to \gamma/Z\, (k^{++}\to \mu^+\mu^+)$; these are not exactly background free, even though invariant mass distributions and angular observables can be used to isolate new-physics contributions.
The subleading channels 
$\mu^+\mu^+\to \mu^+\tau^+$ and $\mu^+\mu^+\to \gamma/Z\, (k^{++}\to \mu^+\tau^+)$
on the other hand are smoking-gun observables.

The texture from Eq.~\eqref{eq:ZeeBabu_texture} does not induce any interesting signatures in the $\mu^+ e^-$ run, but other textures might, see Fig.~\ref{fig:doublycharged}(c). For example, a $\mu\mu$ and $ee$ entry in $g$ would give the very clean $\mu^+e^-\to \mu^- e^+ $
(in addition to $\mu^+\mu^+\to e^+ e^+$), allowed by current muonium-conversion constraints, as shown in Fig.~\ref{fig:ZeeBabuplot}.

We also show other relevant constraints in Fig.~\ref{fig:ZeeBabuplot}. The $(g-2)_\mu$ excluded region is shown by the black shaded region on top left corner. The vertical pink shaded region is the current LHC bound~\cite{ATLAS:2022pbd}, and the vertical pink dashed line is the future HL-LHC sensitivity~\cite{Ruiz:2022sct}. Thus, we find that $\mu$TRISTAN will probe a wide range of the Zee--Babu model parameter space well beyond the HL-LHC sensitivity. Similar sensitivities are also achievable at a future $\mu^+\mu^-$ collider~\cite{Chowdhury:2023imd}.  

\subsection{Cocktail model}
The cocktail model~\cite{Gustafsson:2012vj} is an SM extension by two $SU(2)_L$-singlet scalars $h^-$ and $k^{++}$, as well as a second Higgs doublet $H_2$. The field content is reminiscent of the Zee and Zee--Babu models, but here an extra $\mathbb{Z}_2$ symmetry is imposed under which $h^-$ and $H_2$ are odd, which leaves the following relevant terms in the Lagrangian:
\begin{align}
\begin{split}
-\L & \supset 
g\bar{\ell}^c  \ell\, k^{++} 
+ \mu h^- h^- k^{++}
+\kappa \tilde{H}_2^\dagger H_1 h^-\\
&\quad +\xi \tilde{H}_2^\dagger H_1 h^+ k^{--}
+\frac{\lambda_5}{2} (H_1^\dagger H_2)^2
+\hc \,,
\end{split}
\end{align}
where $g$ is once again a symmetric Yukawa matrix in flavor space. Lepton number is broken explicitly if all the above couplings are non-zero.
We assume parameters in the scalar potential so that $\langle H_2\rangle = 0$, leaving the  $\mathbb{Z}_2$ unbroken. In that case, Majorana neutrino masses arise at three-loop level:
\begin{align}
M^\nu \simeq \frac{F_\text{cocktail}}{(16\pi^2)^3\,m_{k^{++}}}\,  M_\ell g M_\ell \,,
\end{align}
where $F_\text{cocktail}$ is a complicated dimensionless loop function that depends on scalar masses and couplings~\cite{Geng:2014gua,Cepedello:2020lul}.
The three-loop suppression factor and additional suppression by charged-lepton masses require large entries in $g$ that are easily in the non-perturbative regime, even when all scalar masses are close to their experimental limits and the scalar-potential couplings as large as allowed by perturbative unitarity.
To keep $g$ perturbative and evade stringent constraints from muon LFV, one is more or less forced to consider the two-zero texture $A_1$ for $M^\nu$~\cite{Geng:2014gua,Cepedello:2020lul}, which then results in a $g$ matrix
\begin{align}
g &\propto \matrixx{
0 & 0 & 1 \\
0 & \frac{m_e m_\tau}{m_\mu^2}\, \frac{M^\nu_{\mu\mu}}{M^\nu_{e\tau}} & \frac{m_e}{m_\mu}\, \frac{M^\nu_{\mu\tau}}{M^\nu_{e\tau}} \\
1 & \frac{m_e}{m_\mu}\, \frac{M^\nu_{\mu\tau}}{M^\nu_{e\tau}} & \frac{m_e}{m_\tau}\, \frac{M^\nu_{\tau\tau}}{M^\nu_{e\tau}}
}
\sim\matrixx{
0 & 0 & 1 \\
0 & 0.24 & 0.01 \\
1 & 0.01 & 6\times 10^{-4}} 
\label{eq:Cocktail_texture}
\end{align}
and the neutrino-parameter predictions from the first row of Tab.~\ref{tab:texture_zeros}.
The strongest LFV constraint mediated by $k^{++}$ then comes from $\tau^-\to e^+ \mu^-\mu^-$, requiring $|g_{e\tau}| < 0.17 \, m_{k^{++}}/\unit{TeV}$, although, by coincidence,  $\mu\to e\gamma$ gives essentially the same limit for this texture.

The LFV constraints of this texture are severe enough that $\mu$TRISTAN in the $\mu^+ e^-$ mode would not observe the characteristic $\mu^+ e^-\to \tau^+ \mu^-$, see Fig.~\ref{fig:ZeeBabuplot}.
However, $\mu$TRISTAN in the $\mu^+\mu^+$ run could potentially see $\mu^+\mu^+\to e^+ \tau^+$ or $\mu^+\mu^+\to k^{++} \gamma/Z$ followed by prompt $k^{++}\to e^+\tau^+$ decays. 

Notice that the $\mathbb{Z}_2$ symmetry renders the lightest particle among the $H_2$ and $h^-$ stable. We can choose scalar-potential parameters to make this one of the neutral scalars inside $H_2$, which could then form dark matter. We will not discuss this here since there is very limited connection to $\mu$TRISTAN.

\subsection{Type-II or triplet seesaw}

In the type-II or triplet seesaw mechanism~\cite{Konetschny:1977bn,Magg:1980ut,Schechter:1980gr,Cheng:1980qt,Mohapatra:1980yp}, we extend the SM by an $SU(2)_L$-triplet with hypercharge $+2$, usually written as the $SU(2)_L$ matrix
\begin{align}
\Delta = \begin{pmatrix}
\Delta^+/\sqrt{2} & \Delta^{++}\\
\Delta^0 & -\Delta^+/\sqrt{2}
\end{pmatrix} .
\end{align}
This triplet couples to the left-handed lepton doublets $L_{e,\mu,\tau}$ and the SM scalar doublet $H$, giving rise to the Lagrangian
\begin{align}
-\L \supset  Y \bar{L}^c \ii\sigma_2 \Delta L +\mu H^\dagger \ii\sigma_2 \Delta H^*+ \hc 
\end{align}
This Lagrangian breaks lepton number and induces a small vacuum expectation value $\langle \Delta^0\rangle = v_\Delta/\sqrt{2}$, which in turn generates the Majorana neutrino mass matrix $M^\nu = \sqrt{2} Y v_\Delta$. The Yukawa couplings thus inherit the structure from the neutrino mass matrix but come with an unknown scaling factor $v_\Delta$. 

In the limit of $v_\Delta \ll v$, the mass eigenstates that dominantly come from the triplet, $H^{++}\simeq \Delta^{++}$, $H^+\simeq \Delta^{+}$, $H\simeq\sqrt2 \re\,\Delta^0$, and $A\simeq \sqrt2\im\,\Delta^0$, have mass splittings
\begin{align}
m_H^2 \simeq m_A^2 \simeq m_{H^+}^2+\frac{\lambda_4 v^2}{4}\simeq m_{H^{++}}^2+\frac{\lambda_4 v^2}{2} \,,
\label{eq:splittings}
\end{align}
specified exclusively by the coupling $\lambda_4\, H^\dagger \Delta \Delta^\dagger H$~\cite{Arhrib:2011uy, Mandal:2022zmy}.
For simplicity we will assume an almost degenerate spectrum here, even though a mass splitting could resolve~\cite{Kanemura:2022ahw,Heeck:2022fvl,Butterworth:2022dkt} the recently observed discrepancy in CDF's $W$-boson mass measurement~\cite{CDF:2022hxs}.
The large Yukawa couplings required to produce $\Delta^{++}$ at $\mu$TRISTAN also lead to strong constraints from searches at the LHC, which exclude masses below 1 TeV~\cite{ATLAS:2022pbd} and can be improved at the HL-LHC~\cite{Ashanujjaman:2021txz}.

Even more importantly, the triplet scalars induce LFV decays, for example~\cite{Pich:1984uoh,Akeroyd:2009nu,Dev:2018sel,Mandal:2022zmy}
\begin{align}
&\BR (\mu\to e\gamma)\simeq \frac{\alpha_\text{EM} \left|(M^{\nu\,\dagger} M^\nu)_{e\mu}\right|^2}{48\pi G_F^2 v_\Delta^4}\left(\frac{1}{m_{H^{+}}^2}+\frac{8}{m_{H^{++}}^2}\right)^2,\\
&\BR (\mu^+\to e^+e^-e^+)\simeq 4\frac{\left|M^\nu_{ee} M^\nu_{\mu e}\right|^2}{G_F^2 v_\Delta^4 m_{H^{++}}^4}\,, 
\end{align}
where $G_F$ is the Fermi coupling constant. 
$\mu\to e\gamma$ is particularly important because the prefactor  $ \left|(M^{\nu\,\dagger} M^\nu)_{e\mu}\right|^2$ is completely specified by the known neutrino oscillation parameters~\cite{Chakrabortty:2012vp} and is limited from below by $(\unit[0.016]{eV})^4$, using the $2\sigma$ range from {\tt NuFit 5.2}~\cite{Esteban:2020cvm}. The current limit $\BR (\mu\to e\gamma)<4.2\times 10^{-13}$~\cite{MEG:2016leq} then gives $m_{\Delta^{++}}> \unit[1.5]{TeV}(\unit{eV}/v_\Delta)$. The $\mu\to e\gamma$ limit can be improved by almost an order of magnitude with MEG-II~\cite{Baldini:2013ke,Meucci:2022qbh} but will eventually be surpassed by muon-conversion in Mu2e~\cite{Mu2e:2014fns,Mu2e-II:2022blh}, which probes the same coupling in our case and effectively has a sensitivity down to $\BR (\mu\to e\gamma)<2\times 10^{-14}$. This would improve the limit to $m_{\Delta^{++}}> \unit[3]{TeV}(\unit{eV}/v_\Delta)$.

Notice that the other LFV decays, notably $\mu\to 3 e$~\cite{SINDRUM:1987nra}, could give even stronger limits on $ v_\Delta m_{\Delta^{++}}$, especially with the upcoming Mu3e~\cite{Mu3e:2020gyw}, but depend on the so-far unknown neutrino parameters such as the lightest neutrino mass and the Majorana CP phases. These allow us, for example, to set $M^\nu_{ee} = 0$ and thus eliminate $\mu\to 3e$ entirely. For simplicity we will therefore ignore these other LFV processes and only consider the unavoidable $\mu\to e \gamma$.

\begin{figure}[!t]
    \centering
    \includegraphics[width=0.46\textwidth]{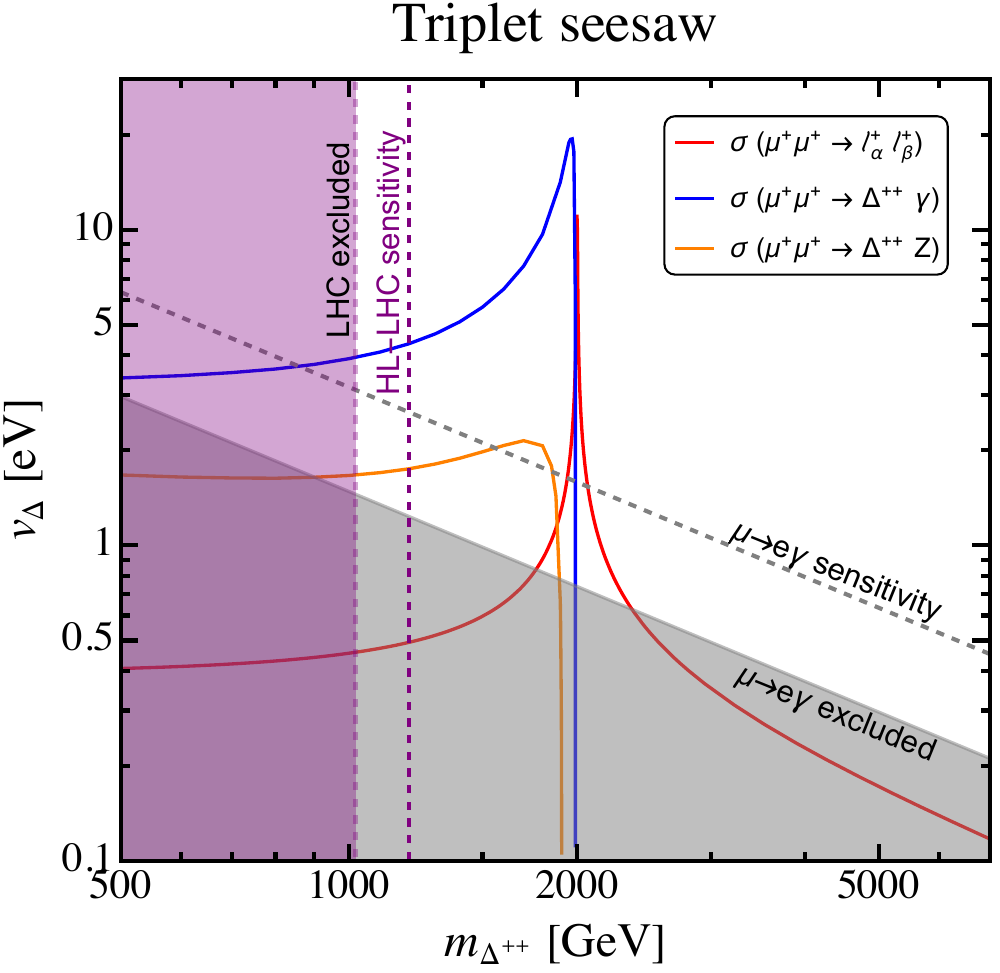}
    \caption{ 
    $\mu$TRISTAN sensitivity to the triplet/type-II seesaw model parameter space for various channels as shown in Fig.~\ref{fig:doublycharged}. We have set $M^\nu_{\mu\mu} = \unit[0.05]{eV}$ to fix the $\Delta^{++}\mu\mu$ coupling, see text for details.
    }
    \label{fig:typeIIplot}
\end{figure}

In Fig.~\ref{fig:typeIIplot}, we show the LFV and LHC constraints together with the $\mu$TRISTAN sensitivities in various channels. We have implemented the model file in {\tt FeynRules}~\cite{Alloul:2013bka}  and computed the cross sections using {\tt MadGraph5\_aMC@NLO}~\cite{Alwall:2014hca}. To specify the production Yukawa coupling $Y_{\mu\mu}$ we set $M^\nu_{\mu\mu} =\unit[0.05]{eV}$; this satisfies the cosmology bound  $\sum m_\nu < \unit[0.12]{eV}$~\cite{Planck:2018vyg}, otherwise we could go to larger  $M^\nu_{\mu\mu} $ values and increase the $\mu$TRISTAN cross sections without changing the LFV bound.

The cross section $\sigma (\mu^+\mu^+\to \ell^+_\alpha \ell^+_\beta)$ scales with $|M^\nu_{\mu\mu}|^2|M^\nu_{\alpha\beta}|^2$, at least away from the resonance. 
The on-shell produced $\Delta^{++}$ has decay rates into charged leptons proportional to $|M^\nu_{\alpha\beta}|^2$. Our current lack of information about the lightest neutrino mass and the CP phases preclude us from making definite predictions for these final states, but this will improve with future neutrino data~\cite{Ellis:2020hus}. Generically, we expect final states with more muons and tauons than electrons at $\mu$TRISTAN from $\Delta^{++}$ processes for normal-ordered neutrino masses. Diboson decays $\Delta^{++}\to W^+ W^+$ are heavily suppressed by $v_\Delta$ in our region of interest~\cite{FileviezPerez:2008jbu,Rodejohann:2010jh,Rodejohann:2010bv,Melfo:2011nx}. Similarly, the cascade decays of $\Delta^{++}$ involving neutral or singly-charged scalars depend on the choice of mass spectrum and can be ignored here.

Unlike for the doubly charged scalars in the Zee--Babu or cocktail models, the $\Delta^{++}$ in the triplet model cannot generate clean $\mu^+ e^-\to \ell^+ \ell'^-$ signatures in $\mu$TRISTAN's first run, since this region of parameter space is already excluded by $\mu\to e\gamma$ (Fig.~\ref{fig:typeIIplot}).

\subsection{Other neutrino mass models}

The $\mu^+\mu^+$ mode of $\mu$TRISTAN will also be uniquely sensitive to the LNV/LFV signatures arising from other neutrino mass models. For instance, the heavy neutral leptons appearing in type-I~\cite{Minkowski:1977sc, Mohapatra:1979ia, Gell-Mann:1979vob, Yanagida:1979as, Glashow:1979nm} and type-III~\cite{Foot:1988aq} seesaw models will induce a clean LNV signal $\mu^+\mu^+\to W^+W^+\to {\rm jets}$, which is like an inverse neutrinoless double beta decay $e^-e^-\to W^-W^-$~\cite{Rizzo:1982kn,Gluza:1995ix,Belanger:1995nh,Ananthanarayan:1995cn} but in the muon sector~\cite{Heusch:1995yw}. This channel has been recently analyzed in Refs.~\cite{Yang:2023ojm, Jiang:2023mte}, so we will not repeat this analysis here. Similarly, the $\mu$TRISTAN sensitivities for the neutral and/or doubly-charged scalars derived here can also be applied to other models, such as the left--right symmetric model~\cite{Mohapatra:1974gc, Mohapatra:1974hk, Senjanovic:1975rk}, and other radiative neutrino mass models~\cite{Cai:2017jrq}, although the connection to neutrino mass may not be as direct as in the models studied here.

\section{Conclusion}
\label{sec:conclusions}

Neutrino masses provide the most convincing laboratory evidence for physics beyond the SM, making searches for the underlying new particles highly motivated.
In this article, we have shown that $\mu^+ e^-$ and $\mu^+\mu^+$ colliders in the vein of the recently proposed $\mu$TRISTAN experiment offer a new way to search for a variety of neutrino mass models. As exemplified by several benchmark scenarios of the popular Zee, Zee--Babu, cocktail, and triplet seesaw models, we showed that $\mu$TRISTAN could probe regions of parameter space that are out of reach of other experiments, be it future hadron colliders or future low-energy LFV searches.

\section*{Acknowledgements}
The work of BD is supported in part by the U.S. Department of Energy under grant No. DE-SC 0017987 and by a URA VSP fellowship. The work of JH and AT was supported in part by the National Science Foundation under Grant PHY-2210428. For facilitating portions of this research, BD and AT wish to acknowledge the Center for Theoretical Underground Physics and Related Areas (CETUP*), The Institute for Underground Science at Sanford Underground Research Facility (SURF), and the South Dakota Science and Technology Authority for hospitality and financial support, as well as for providing a stimulating environment.
BD and JH would like to thank the Fermilab Theory Group for their hospitality during the completion of this work.

\bibliographystyle{utcaps_mod}
\bibliography{BIB.bib}

\end{document}